# Automated pharyngeal phase detection and bolus localization in videofluoroscopic swallowing study: Killing two birds with one stone?

Andrea Bandini[1,2,3], Sana Smaoui[1,4], Catriona M. Steele[1,4] *

[1] KITE Research Institute – Toronto Rehabilitation Institute, University Health Network, Toronto, ON, M5G 2A2, Canada

[2] The BioRobotics Institute, Scuola Superiore Sant'Anna, Pisa, Italy

[3] Department of Excellence in Robotics and AI, Scuola Superiore Sant'Anna, Pisa, Italy

[4] Rehabilitation Sciences Institute, Temerty Faculty of Medicine, University of Toronto, Ontario, Canada

*Corresponding author (email: catriona.steele@uhn.ca)

## ABSTRACT

**Background and Objective**: The videofluoroscopic swallowing study (VFSS) is a gold-standard imaging technique for assessing swallowing, but analysis and rating of VFSS recordings is time consuming and requires specialized training and expertise. Researchers have recently demonstrated that it is possible to automatically detect the pharyngeal phase of swallowing and to localize the bolus in VFSS recordings via computer vision approaches, fostering the development of novel techniques for automatic VFSS analysis. However, training of algorithms to perform these tasks requires large amounts of annotated data that are seldom available. In this paper, we demonstrate that the challenges of pharyngeal phase detection and bolus localization can be solved together using a single approach.

**Methods**: We propose a deep-learning framework that jointly tackles pharyngeal phase detection and bolus localization in a weakly-supervised manner, requiring only the initial and final frames of the pharyngeal phase as ground truth annotations for the training. Our approach stems from the observation that bolus presence in the pharynx is the most prominent visual feature upon which to infer whether individual VFSS frames belong to the pharyngeal phase. We conducted extensive experiments with multiple convolutional neural networks (CNNs) on a dataset of 1245 bolus-level clips from 59 healthy subjects.

**Results**: We demonstrated that the pharyngeal phase can be detected with an F1-score higher than 0.9. Moreover, by processing the class activation maps of the CNNs, we were able to localize the bolus with promising results, obtaining correlations with ground truth trajectories higher than 0.9, without any manual annotations of bolus location used for training purposes.

**Conclusions**: Once validated on a larger sample of participants with swallowing disorders, our framework will pave the way for the development of intelligent tools for VFSS analysis to support clinicians in swallowing assessment.

# 1 INTRODUCTION

Dysphagia (swallowing impairment) may occur as a consequence of ageing, neurodegenerative diseases, stroke, and other damage to the central nervous system and/or cranial nerves [1]–[4]. Dysphagia has a negative impact on the quality of life of the affected individuals, and increases the risk of aspiration pneumonia, choking, malnutrition, and dehydration. Timely and accurate instrumental assessment is essential to prevent adverse events and plan appropriate interventions, such as diet texture modifications [5].

The videofluoroscopic swallowing study (VFSS) is a current gold standard clinical assessment for dysphagia [6], [7]. VFSS is an X-ray imaging technique that provides a dynamic view of the anatomical structures involved in the swallow, such as the oropharynx, pharynx, and esophagus (Figure 1, top row). During this exam, patients swallow food and liquid items prepared with a radiocontrast agent (i.e., barium sulfate). Trained speech-language pathologists observe bolus motion and movements of the anatomical structures of interest (e.g., pharynx, hyoid bone, upper esophageal sphincter, etc.) to make clinical decisions and plan interventions.

VFSS is the only technique able to show the bolus throughout the three phases of the swallow, namely: 1) the oral phase - where the food or liquid is being processed and transported through the oral cavity; 2) the pharyngeal phase - where a series of coordinated movements transport the bolus through the pharynx and upper esophageal sphincter (UES) while simultaneously protecting the entrance to the lower airway from penetration-aspiration; and 3) the esophageal phase - when the bolus passes through the esophagus and travels towards the stomach [8], [9]. In particular, the pharyngeal phase is the time window of the swallow event that contains most of the clinical parameters of interest, and a thorough evaluation of it is essential for detecting the presence of penetration-aspiration [10].

In order to standardize the VFSS assessment and plan optimal interventions, several video rating procedures have been proposed [10]–[13]. However, VFSS rating remains a lengthy process, as it requires frame-by-frame inspection of the recorded video [10]. Moreover, raters require specialized training that needs to be constantly refreshed to maintain high standards of inter-rater agreement. Thus, it would be desirable if some parts of the VFSS rating could be automatized, to support clinicians in assessment and ease clinical workload.

Several artificial intelligence approaches for VFSS analysis have been proposed over the past few years [7], [9], [14]–[19]. Of particular interest are those that tried either to detect the pharyngeal phase [9], [15] or to locate the bolus during the swallow [16]. The reliable detection of temporal events (i.e., the pharyngeal phase) and spatial information (i.e., bolus localization) would constitute an important aid to clinicians during VFSS rating, as this would expedite the rating of VFSS recordings, remove inter subject variability, and - especially in case of bolus localization - support the identification of aspiration-penetration events. From an algorithmic perspective, these two problems have always been tackled independently, for example: the pharyngeal phase has usually been detected in a video segmentation fashion using convolutional neural networks (CNNs) [9], [14], [15], whereas the bolus has been localized via image segmentation algorithms [16]. In particular, the latter task requires a large amount of pixel-level training data, that are seldom available due to the lack of public datasets and the huge amount of time required to conduct the annotations.

Building upon the existing literature and the need for automated approaches for VFSS analysis, we propose a deep-learning framework that tackles pharyngeal phase detection and bolus localization simultaneously in a weakly-supervised manner, requiring only the initial and final frames of the pharyngeal phase as ground truth annotations for training. Our approach stems from the observation that bolus presence in the pharynx is the most prominent visual characteristic upon which to infer whether a VFSS frame belongs to the pharyngeal phase or not. Our hypothesis is that a CNN trained to detect the pharyngeal phase [15] must rely on the presence of the bolus in the pharynx to make predictions, and that its class activation maps (e.g., extracted using Grad-CAM [20]) should exhibit higher "attention" around the bolus region, providing a coarse bolus location that can be refined via further image processing (see Figure 1 and supplementary material).

Specifically, the main contributions of this manuscript are to:

- Train and compare multiple CNN architectures for detecting the pharyngeal phase and understand how far we are from implementing these algorithms in clinical practice, by comparing classification performance with the inter-rater reliability of trained annotators.

- Propose a weakly-supervised bolus localization method that requires no supervision in terms of bolus coordinates or pixel-level segmentation masks, but only the beginning and the end of the pharyngeal phase.

- Test the robustness of pharyngeal phase detection and bolus localization across different consistencies, to understand whether there is any relationship between the performance of the automated approach and liquid bolus consistency.

## 1.1 Related work

### 1.1.1 Computer vision for VFSS analysis

Several artificial intelligence applications in swallowing science have been proposed over the past few years, fostering novel approaches aimed at improving and automating swallowing and dysphagia assessment [7]. Most of these approaches involve the use of computer vision - for example, to analyze VFSS recordings [9], [14]–[17], [19], [21]–[23] - or signal processing techniques to extract and analyze acoustic, accelerometric, and/or electromyographic signals [24]–[30]. Despite emerging evidence that wearable sensors may be able to detect key outcomes when VFSS is not available [26], [27], [31], VFSS still remains the most widely accepted imaging technique for assessing dysphagia [7].

The importance of VFSS has stimulated many researchers to develop automated approaches to support clinicians in the interpretation of the recorded video, for example, by detecting structures and anatomical points of interest [18], [22], [32], tracking hyoid bone movements [19], [21], [23], [33]–[35], segmenting the bolus during swallow events [16], and detecting the pharyngeal phase from the raw videos [9], [14], [15]. The advent of deep learning has considerably advanced the state of the art of this field. In fact, approaches for tracking hyoid bone movement, which were semi-automatic until a few years ago [23], [33]–[35], have recently became fully automated with improved detection accuracies [19], [21]. This was possible due to the availability of powerful object detection neural networks such as Faster R-CNN [36] and YOLO [37]. Likewise, temporal analysis of VFSS has benefited from the development of powerful 3DCNNs [38] that allowed detection of the pharyngeal phase in VFSS recordings [9], [14]. More recently, Lee et al. [15] demonstrated that comparable performance in detecting the pharyngeal phase could be obtained with 2DCNNs [39], with a significant reduction of computational complexity compared to 3DCNNs.

### 1.1.2 Pharyngeal phase detection and bolus localization: killing two birds with one stone?

Recent results obtained by Lee et. al [9], [15] demonstrated that this phase can be detected with accuracy higher than 90%. Detecting the pharyngeal phase in VFSS recordings involves conducting a binary classification task to determine whether or not each frame contains the bolus within the pharynx. Thus, our hypothesis is that a CNN trained to discriminate pharyngeal and non-pharyngeal frames must rely on bolus presence in the pharynx to make its predictions, as this is the most prominent visual feature for detecting this phase. Hence, the class activation maps obtained when the frames are classified as belonging to the pharyngeal phase, should exhibit higher network "attention" around the bolus' location (see Figure 1 and supplementary material).

To verify our hypothesis, we capitalize on the availability of algorithms that provide visual explanations to CNN predictions, such as Grad-CAM [20], [40]. Since its publication in 2017, Grad-CAM has had a huge impact in computer vision applications, as it can highlight the regions of the image where the CNN is looking to perform its predictions [20], [40]. In our case, this translates into the possibility of performing weakly-supervised bolus localization, where the bolus is detected by using only the predicted class of a frame (i.e., pharyngeal phase).

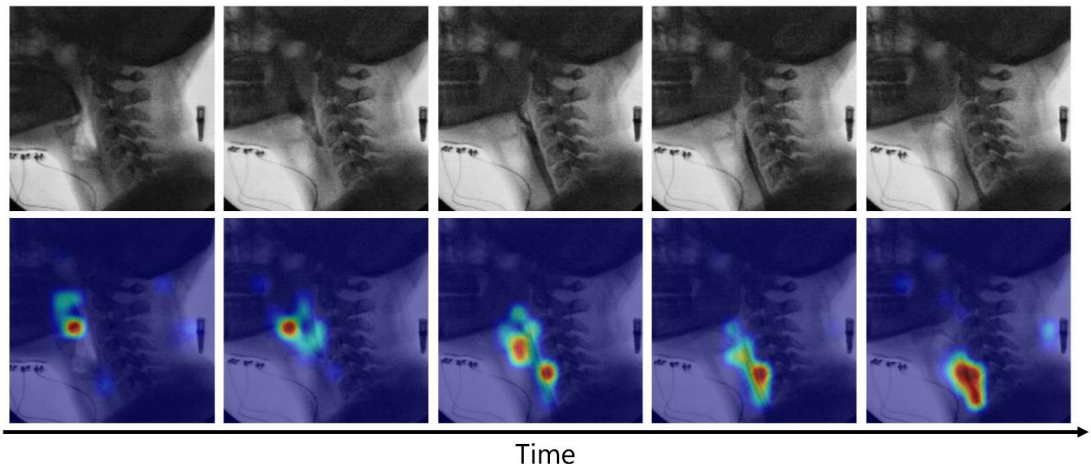

Time

**Figure 1. Heatmaps obtained with Grad-CAM (bottom row) on the input VFSS frames (top row) during the swallow of a bolus. Warmer colors correspond to parts of the images where the CNN attention goes to classify the VFSS frames as belonging to the pharyngeal phase or not. Examples obtained with CNN4 (see Section 2.4).**

# 2 METHODS

## 2.1 Data collection

Seventy-eight participants were recruited for this study. All participants reported no history of swallowing, motor speech, gastroesophageal, or neurological disorders. The complete list of inclusion and exclusion criteria, as well as the details of the acquisition protocol are described in [10]. Briefly, each participant self-administered 27 boluses of low-concentration barium stimuli (20% w/v), taking comfortable sips or spoonfuls. Stimuli were prepared mixing bottled water and powdered barium sulfate (Bracco Diagnostics E-Z-PAQUE, 96% w/w) in five consistencies: thin, slightly thick, mildly thick, moderately thick and extremely thick, according to the definitions of the International Dysphagia Diet Standardisation Initiative (IDDSI) [5]. With the exception of thin boluses, two types of thickeners were used: Nestlé Resource® ThickenUp® Clear and Resource® ThickenUp®. Three boluses for each combination of consistency/thickener were swallowed by each participant [10]. VFSS recordings were conducted in lateral projection and recorded on a KayPENTAX Digital Swallow Workstation at 720×480 pixels resolution and 30 frames per second. All 27 bolus administrations were stored onto a single video recording. The study was approved by the Research Ethics Board at UHN - Toronto Rehabilitation Institute. All participants signed informed consent according to the requirements of the Declaration of Helsinki.

## 2.2 Rating procedure and dataset

The full-length VFSS recordings were split into separate clips, each one containing the administration of a bolus (bolus-level clips). This task was conducted manually, identifying the frames associated with onset and offset of the X-ray, as per ASPEKT guidelines [10]. The bolus-level clips were randomly assigned to pairs of raters from a team of 8 trained speech-language pathologists, who followed the ASPEKT protocol for rating VFSS clips [10]. From the entire set of ASPEKT parameters, in this study, we considered only the time points that delimit the pharyngeal phase, namely the frame of the bolus passing the mandible (BPM - beginning of the pharyngeal phase) and frame of upper esophageal sphincter closure (UESC - end of the pharyngeal phase). The BPM frame is "*the first frame where the leading edge of the bolus touches or crosses the shadow of the ramus of the mandible*", whereas the UESC frame is defined as "*the first frame where the UES achieves closure behind the bolus tail*" [10]. Discrepancies were resolved via consensus meetings with a third rater. The resolved values of BPM and UESC frames were used as the ground truth for developing the deep learning methods to detect the pharyngeal phase (Sec. 2.4).

In this study, we only considered video-clips with single-swallow boluses, namely those with an initial bolus swallow followed by no other events [10]. Moreover, we excluded video-clips whose BPM and UESC frames were considered unratable due to occluded images or poor video quality. The dataset used for the experiments was composed of 1245 bolus-level clips from 59 participants. Inter-rater agreement was calculated on the pre-consensus values using the Pearson's correlation coefficient (r) and the percentage of video clips for which disagreement in frame selection was below three frames (P3). We used a threshold of three frames since this value is commonly used in VFSS rating to decide when a discrepancy needs to be resolved with a third rater [10], [41]. An overview of the dataset's characteristics is reported in Table 1.

**Table 1. Dataset characteristics**

| | | |
|---|---|---|
| Participants (female) | | 59 (33) |
| Age (years) | M ± SD | 44.7 ± 17.9 |
| | Range | 21 - 82 |
| Number of video-clips (n=1245) per bolus consistency | Extremely thick | 251 |
| | Moderately thick | 282 |
| | Mildly thick | 285 |
| | Slightly thick | 299 |
| | Thin | 128 |
| Inter-rater agreement (on 1245 videos) | BPM frame | r = 0.951 |
| | | P3 = 89.08% |
| | UESC frame | r = 0.996 |
| | | P3 = 92.96% |

### 2.3 Data preparation

Each bolus-level clip was split into separate frames, obtaining 185,025 grayscale frames. The central region of interest of each frame was automatically cropped by selecting a square of size 341×341 pixels from the center of the frame. After a preliminary visual inspection of the dataset, we determined that this size of was a good trade-off between keeping the useful anatomical information within the X-ray view and disregarding most of the background part of the frames (see

Figure 2). Afterwards, image contrast was improved with a contrast-limited adaptive histogram equalization as proposed by Wilhelm et al. [42].

To automatically detect the pharyngeal phase, the pre-processed frames were assigned to two classes: class P (i.e., pharyngeal) if the frames were between the BPM and UESC frames, included, and class N (i.e., non-pharyngeal), if the frames were outside the BPM-UESC interval[1]. The entire dataset was randomly split into three subsets, assigning all bolus-level clips of a single participant to one of the following sets: 1) training-set - composed of 752 video-clips (77 thin, 183 slightly thick, 172 mildly thick, 167 moderately thick, 153 extremely thick) from 38 participants (109,203 frames in total); 2) validation-set - composed of 201 video-clips (22 thin, 49 slightly thick, 44 mildly thick, 47 moderately thick, 39 extremely thick) from 9 participants (29,491 frames in total); and 3) test-set - composed of 292 video-clips (29 thin, 67 slightly thick, 69 mildly thick, 68 moderately thick, 59 extremely thick) from 12 participants (47,331 frames in total).

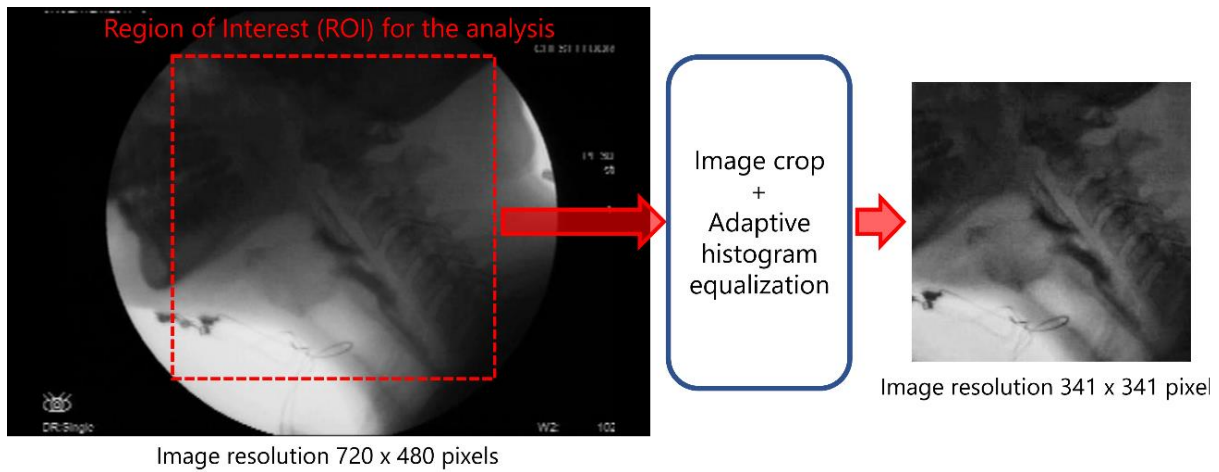

**Figure 2. Overview of the pre-preprocessing steps.**

### 2.4 Supervised pharyngeal phase detection

Following the results obtained by Lee et al. [15], we detected the pharyngeal phase in a frame-by-frame manner using a CNN to predict whether each frame of the bolus-level clips belonged to the pharyngeal phase. We implemented and compared the following backbone architectures: VGG16 [39], InceptionV3 [43], and ResNet50 [44]. Moreover, we designed two architectures composed of 3 and 4 blocks - hereafter CNN3 and CNN4, respectively - where each block contained two stacked convolutional layers with 3×3 kernel size and one max pooling layer with pooling size 2×2. The number of convolutional filters were: 4 for the first block, 8 for the second block; 16 for the third block; and 32 for the fourth block (only in CNN4). Given the binary nature of the pharyngeal phase detection, all models had a softmax layer with two output nodes for the classification and for VGG16, CNN3, and CNN4, the softmax layer was preceded by two fully connected layers of 128 and 64 nodes. Each model was trained from scratch for 100 epochs using Adam as optimizer and batch size equal to 8. For InceptionV3, ResNet50, CNN3, and CNN4 we used an initial learning rate = 1e-3 decayed every 5 epochs, whereas for VGG16 we used an initial learning rate = 1e-5 decayed every 10 epochs. Additional details about the CNN architectures and hyperparameter tuning is reported in the supplementary material.

---

[1] Class N included frames from both the oral and esophageal phase.

For each bolus-level clip of the test set, we identified the BPM frame as the first frame with predicted class equal to P that was followed by at least 3 consecutive P frames. Similarly, we determined the UESC frame as the last frame of predicted class equal to P, which was preceded by at least 3 consecutive frames of class P. Model performance was compared in terms of F1-score and the percentage of frames for which the events of interest were identified with an error less than or equal to 3 frames (P3$_{BPM}$ and P3$_{UESC}$).

## 2.5 Weakly-supervised bolus localization

At test time, we implemented a weakly-supervised approach to estimate bolus location during the pharyngeal phase. Specifically, we applied the Grad-CAM algorithm onto the 5 trained CNN models to extract the class activation maps of the test frames [20], [40]. The coarse bolus location obtained with Grad-CAM was then refined via additional image processing, i.e. image thresholding and morphological operators [45] followed by active contours [46], [47].

### 2.5.1 Coarse localization maps

To obtain the coarse bolus localization maps, we used the original implementation of Grad-CAM proposed by Selvaraju et al., [20], [40] that we briefly summarize here. First, we computed the gradient of the top predicted class with respect to the output feature maps of the last convolutional layer of the CNN. Since the computed gradient has three dimensions (i.e., same shape of the last convolutional layer), a global average pooling was performed to obtain the importance of each channel of the feature maps with respect to the target class, which was then multiplied with the feature maps themselves. The weighted activation maps were passed through a rectified linear unit (ReLU) activation, averaged over the channel dimension, and normalized between 0 and 1 to obtain a single-channel activation map with the same number of rows and columns of the last convolutional layer. The output sizes of activation maps were 7×7 for VGG16 and ResNet50, 8×8 for InceptionV3, 28×28 for CNN3, and 14×14 for CNN4. Finally, the maps were up sampled to the original frame size (i.e., 341×341). Examples of heatmaps with coarse bolus locations are shown in Figure 1.

### 2.5.2 Refined localization maps

After obtaining the class activation maps and thus the coarse bolus location, we implemented additional image processing steps to refine the localization. Specifically: 1) the up sampled activation maps were binarized using a fixed threshold equal to 0.5 of the maximum heatmap intensity; 2) image dilation [45] was applied to fill small holes within the binary mask and if more than one blob was found, only the largest one was retained; 3) the binary maps were used as masks to seek the k darkest pixels within the grayscale VFSS frames (with k = 100) that, due to the presence of the radiocontrast agent, supposedly belonged to the bolus; 4) the convex hull of the k points was used to initialize a geodesic active contour segmentation algorithm [47], which was ran for 100 iterations; and 5) the centroid of the refined shape was computed (Figure 3). For algorithm evaluation purposes we also extracted the bolus bounding box as the rectangle that included the segmented area.

### 2.5.3 Coordinate transformation and performance evaluation

To test the performance of the proposed bolus localization method, a trained rater manually segmented the bolus in 613 frames (135 extremely thick, 143 moderately thick, 101 mildly thick, 125 slightly thick, and 109 thin) extracted from 25 bolus-level clips (5 clips per consistency) randomly selected from the test set used for the pharyngeal phase detection. The analysis was conducted only on frames within the pharyngeal phase, as the motion of the bolus outside this phase was either non-informative (i.e., before BPM frame) or not visible (i.e., post UESC). Moreover, the anterior-inferior corners of the C2 and C4 vertebrae were annotated by the same rater, in order to obtain a new coordinate system whose vertical (y) axis was defined by the C2-C4 cervical spine, with the horizontal (x) axis perpendicular to C2-C4, passing through C4 [10] (Figure 3). The coordinates of the bolus centroid were mapped into the new coordinate system to reduce the bolus motion to the vertical component of its trajectory, as the y-axis can be approximated as parallel to the pharynx [10] (Figure 3).

The performance of the bolus localization algorithm built on the five CNN models described in Section 2.4 was evaluated using the following metrics:

- ***Pearson's correlation coefficient ($r_y$)*** between the ground truth and the estimated vertical component of the bolus trajectory in the transformed coordinate system.

- ***Root mean square error (RMSE)*** between the ground truth and estimated bolus centroid coordinates, normalized to the C2-C4 Euclidean distance.

- ***F1-score of bounding box detection***. F1-score is a measure of detection accuracy which varies between 0 and 1 in case of perfect detection. To obtain the F1-score, the intersection over union (IoU) between the predicted and ground truth bounding boxes (obtained as the rectangles that included the manually segmented areas) was first calculated. By setting a pre-defined threshold on the IoU, we obtained the number of true positives (TP - frames with IoU $\geq$ threshold), false positives (FP - frames with IoU < threshold), and false negatives (FN - frames with no detection or with IoU < threshold), from which we computed the precision (i.e., TP/TP+FP) and recall (TP/TP+FN). Finally, the F1-score was calculated as 2 * (Recall * Precision) / (Recall + Precision). For each algorithm, the F1-score was obtained across values of IoU threshold between 0.25 and 0.75.

A non-parametric Friedman test was conducted to test for differences between the RMSEs obtained with the five CNN models. In case of significance level of the test (p < .05) a Tukey's honestly significant difference test for multiple comparisons was conducted to identify which model or models produced the lowest errors.

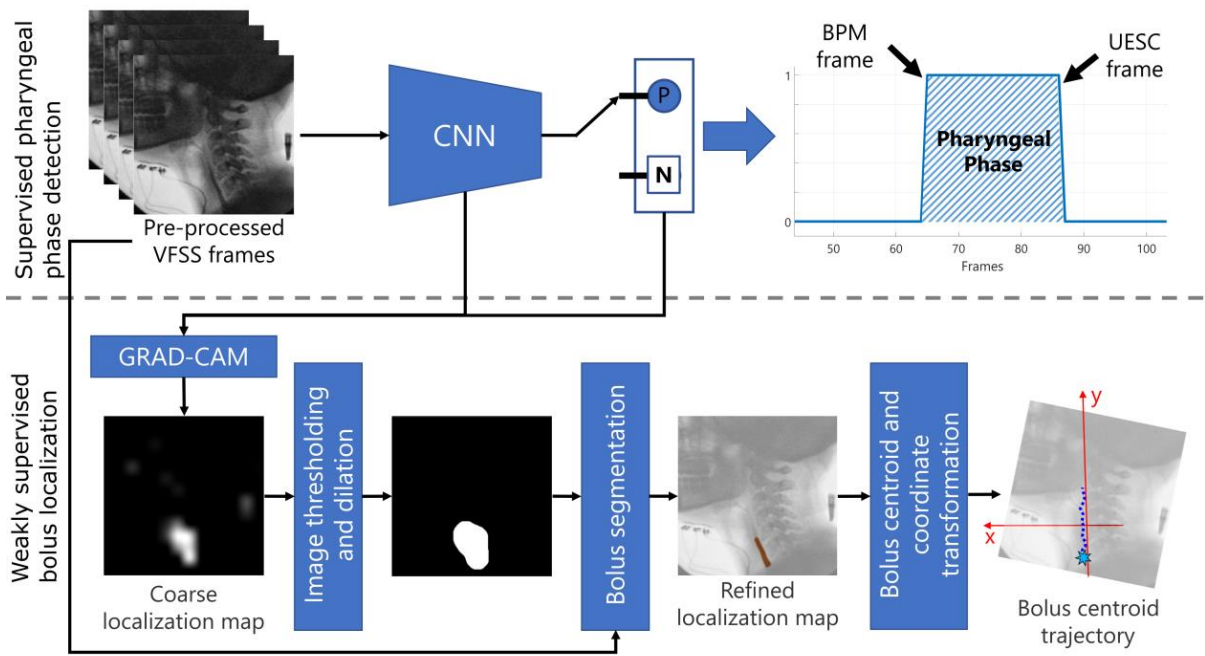

**Figure 3. Processing pipeline used to detect the pharyngeal phase and the bolus location during the swallow in bolus-level clips.**

**Top row (Supervised pharyngeal phase detection) -** Each frame was classified by the CNN as pharyngeal (P) or non-pharyngeal (N). The BPM frame was identified as the first frame with predicted class equal to P that was followed by at least 3 consecutive P frames, whereas the UESC frame was identified as the last frame with predicted class equal to P, which was preceded by at least 3 consecutive frames of class P.

**Bottom row (Weakly supervised bolus localization) -** We refined the coarse localization maps obtained with Grad-CAM via additional image processing steps. Specifically: we binarized the activation maps and used image dilation to fill small holes (only the largest blob was retained). The binary map was used as mask to seek the 100-darkest pixels within the pre-processed VFSS frames, and the convex hull of the 100-darkest pixels was used to initialize a geodesic active contour segmentation algorithm. Finally, the centroid of the refined shape was computed and mapped to a coordinate system with vertical axis defined by the C2-C4 spinal segment.

# 3 RESULTS

## 3.1 Pharyngeal phase detection

In this section, we report the pharyngeal phase detection performance. Beyond the comparison across the five CNN architectures, we compared the results of the automated analysis with the inter-rater agreement of the trained clinical raters. For BPM and UESC frame estimation, we set the P3 values obtained from the pre-consensus ratings as the thresholds to exceed to state that such an automated approach is as good as a trained rater in detecting these time-points. We first report the results obtained on the entire test-set (Sec. 3.1.1) followed by the results per bolus consistency (Sec. 3.1.2).

### 3.1.1 Overall results

The pharyngeal phase detection performance obtained on the entire test set is shown in Table 2. The best results in terms of F1-score were obtained with ResNet50 (F1 = 0.909) closely followed by VGG16 (F1 = 0.906). ResNet50 and VGG16 were also the models that yielded the top two $P3_{BPM}$ and $P3_{UESC}$ values. In terms of comparison with the manual rating, all $P3_{BPM}$ values were below the inter-rater agreement, whereas for the UESC frame, all models but CNN3 produced $P3_{UESC}$ values higher than the inter-rater agreement.

**Table 2. Pharyngeal phase detection performance obtained on the entire test-set. The best results for each metric are highlighted in bold. $P3_{BPM}$ and $P3_{UESC}$ values higher than or equal to the inter-rater agreement are highlighted in green.**

|  | CNN backbone | F1-score | $P3_{BPM}$ (%) | $P3_{UESC}$ (%) |
|---|---|---|---|---|
| CNN models for pharyngeal phase detection | VGG16 | 0.906 | 79.11 | **94.86** |
|  | InceptionV3 | 0.885 | 76.37 | 92.47 |
|  | ResNet50 | **0.909** | **80.82** | 94.52 |
|  | CNN3 | 0.864 | 75.00 | 85.27 |
|  | CNN4 | 0.884 | 75.34 | 91.44 |
| *Inter-rater agreement* |  |  | *86.99* | *89.73* |

### 3.1.2 Results by consistency

The pharyngeal phase detection performance obtained for the five bolus consistencies is reported in Table 3. For all consistencies, ResNet50 and/or VGG16 yielded the best F1-score. In general, it was possible to obtain F1-scores higher than 0.9 for all consistencies except for mildly thick boluses, with maximum F1-score equal to 0.934 produced by ResNet50 on thin boluses.

ResNet50 and VGG16 produced the best results in terms of BPM frame detection, but with $P3_{BPM}$ values always below the trained rater inter-rater agreement. Only for moderately thick boluses, ResNet50 yielded $P3_{BPM}$ equal to the inter-rater agreement. It is also possible to notice an increase in the BPM detection performance with the decrease of bolus thickness, with the best results obtained with slightly thick and thin boluses.

For the UESC frame detection, it was always possible to obtain $P3_{UESC}$ values between 90 and 100%, which in most cases were higher than the inter-rater agreement. Specifically, VGG16 produced the highest $P3_{UESC}$ value in 3 out of 5 consistencies (i.e., mildly, moderately, and extremely thick), whereas InceptionV3 and ResNet50 were the best models in thin and slightly thick boluses, respectively. Moreover, ResNet50 was the only CNN model that exceeded the inter-rater agreement for all consistencies. Unlike BPM frame detection, however, there was no clear trend of performance across consistency.

**Table 3. Pharyngeal phase detection performance by bolus consistency. The best results for each metric are highlighted in bold. P3$_{BPM}$ and P3$_{UESC}$ values higher than or equal to the inter-rater agreement are highlighted in green.**

| | CNN backbone | F1-score | P3$_{BPM}$ (%) | P3$_{UESC}$ (%) |
|---|---|---|---|---|
| Thin boluses | VGG16 | 0.921 | **89.66** | 89.66 |
| | InceptionV3 | 0.927 | **89.66** | **93.10** |
| | ResNet50 | **0.934** | **89.66** | 89.66 |
| | CNN3 | 0.893 | **89.66** | 75.86 |
| | CNN4 | 0.901 | **89.66** | 82.76 |
| *Inter-rater agreement* | | | *93.10* | *86.21* |
| Slightly thick boluses | VGG16 | 0.914 | **92.54** | 94.09 |
| | InceptionV3 | 0.904 | 89.55 | 92.54 |
| | ResNet50 | **0.927** | 89.55 | **100.00** |
| | CNN3 | 0.901 | 86.57 | 86.57 |
| | CNN4 | 0.894 | 88.06 | 91.04 |
| *Inter-rater agreement* | | | *94.03* | *95.52* |
| Mildly thick boluses | VGG16 | **0.881** | 75.36 | **91.30** |
| | InceptionV3 | 0.850 | 75.36 | 88.41 |
| | ResNet50 | 0.869 | **76.81** | 89.86 |
| | CNN3 | 0.848 | 71.01 | 85.51 |
| | CNN4 | 0.868 | 75.36 | 89.86 |
| *Inter-rater agreement* | | | *86.96* | *88.41* |
| Moderately thick boluses | VGG16 | **0.925** | 77.94 | **100.00** |
| | InceptionV3 | 0.913 | 72.06 | 97.06 |
| | ResNet50 | **0.925** | **82.35** | 97.06 |
| | CNN3 | 0.875 | 70.59 | 88.24 |
| | CNN4 | 0.890 | 69.12 | 94.12 |
| *Inter-rater agreement* | | | *82.35* | *88.24* |
| Extremely thick boluses | VGG16 | 0.900 | 64.41 | **96.61** |
| | InceptionV3 | 0.860 | 61.02 | 91.53 |
| | ResNet50 | **0.908** | **69.49** | 93.22 |
| | CNN3 | 0.830 | 64.41 | 84.75 |
| | CNN4 | 0.876 | 61.02 | 94.92 |
| *Inter-rater agreement* | | | *81.36* | *88.14* |

### 3.2 Weakly-supervised bolus localization

In this section, we report the bolus localization performance during the pharyngeal phase. The test-set considered for these analyses was limited to the subset of bolus-level clips with available manual annotation of bolus location (i.e., 613 frames from 25 video clips).

#### 3.2.1 Overall Results

The overall bolus localization performance is reported in Table 4 and Figure 4. ResNet50 produced the highest correlation with the ground truth trajectories ($r_y = 0.934$). Friedman's test showed significant differences in RMSE among the five algorithms ($p < .001$). The lowest RMSE values were produced by CNN3, which were significantly lower than the other four models. The low RMSE values obtained with CNN3 are also reflected in F1-scores higher than the other four models in most of the IoU thresholds (Figure 4). Examples of bolus localization compared with ground truth annotations are shown in Figure 5.

#### 3.2.2. Results by consistency

The bolus localization performance per bolus consistency is reported in Table 5. For thicker consistencies (mildly, moderately, and extremely thick) the highest correlations with the ground truth trajectories were obtained with CNN4, whereas for thin and slightly thick boluses, the highest correlations were obtained with ResNet50. An increase of $r_y$ values with the increase of bolus thickness can be noticed in Figure 6. However, this increase is not visible in ResNet50, which produced consistent performance across the five consistencies, with $r_y$ values always higher than 0.9.

The lowest RMSE values were obtained with CNN3 (all consistencies) and CNN4 (all consistencies except thin boluses). ResNet50 yielded RMSE values that were not statistically different from the CNN3 model in moderately and slightly thick boluses, whereas RMSE values obtained with InceptionV3 were not significantly different from those obtained with CNN3 in thin and slightly thick boluses.

**Table 4. Overall bolus localization performance. Best results for each metric are highlighted in bold. For the RMSE we reported the median and the inter-quartile range (in parentheses).**

|  | CNN backbone | $r_y$ | RMSE |
|---|---|---|---|
| CNN Models | VGG16 | 0.432 | 0.606 (0.400-0.917) |
|  | InceptionV3 | 0.878 | 0.255 (0.146-0.430) |
|  | ResNet50 | **0.934** | 0.306 (0.193-0.460) |
|  | CNN3 | 0.837 | **0.189 (0.099-0.320)** |
|  | CNN4 | 0.886 | 0.230 (0.114-0.390) |
|  | Friedman's test | | $X^2(4) = 583.87$, $p < .001$ |

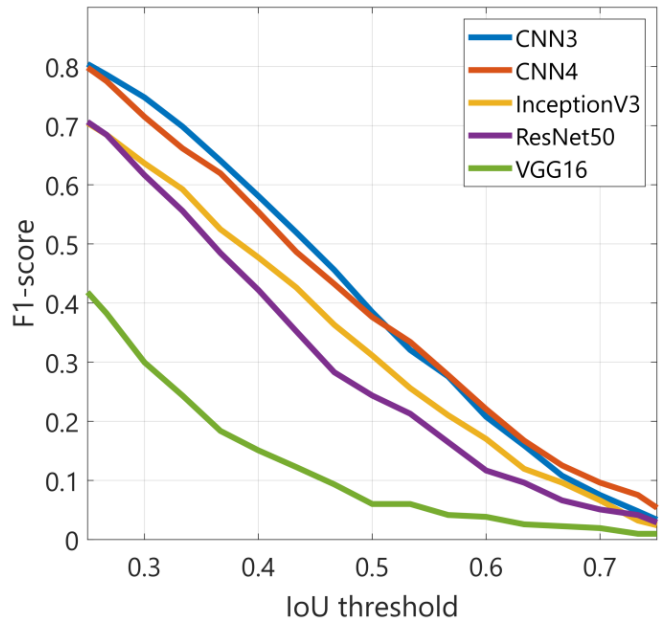

**Figure 4. F1-scores of bolus localization obtained by changing the IoU thresholds between 0.25 and 0.75.**

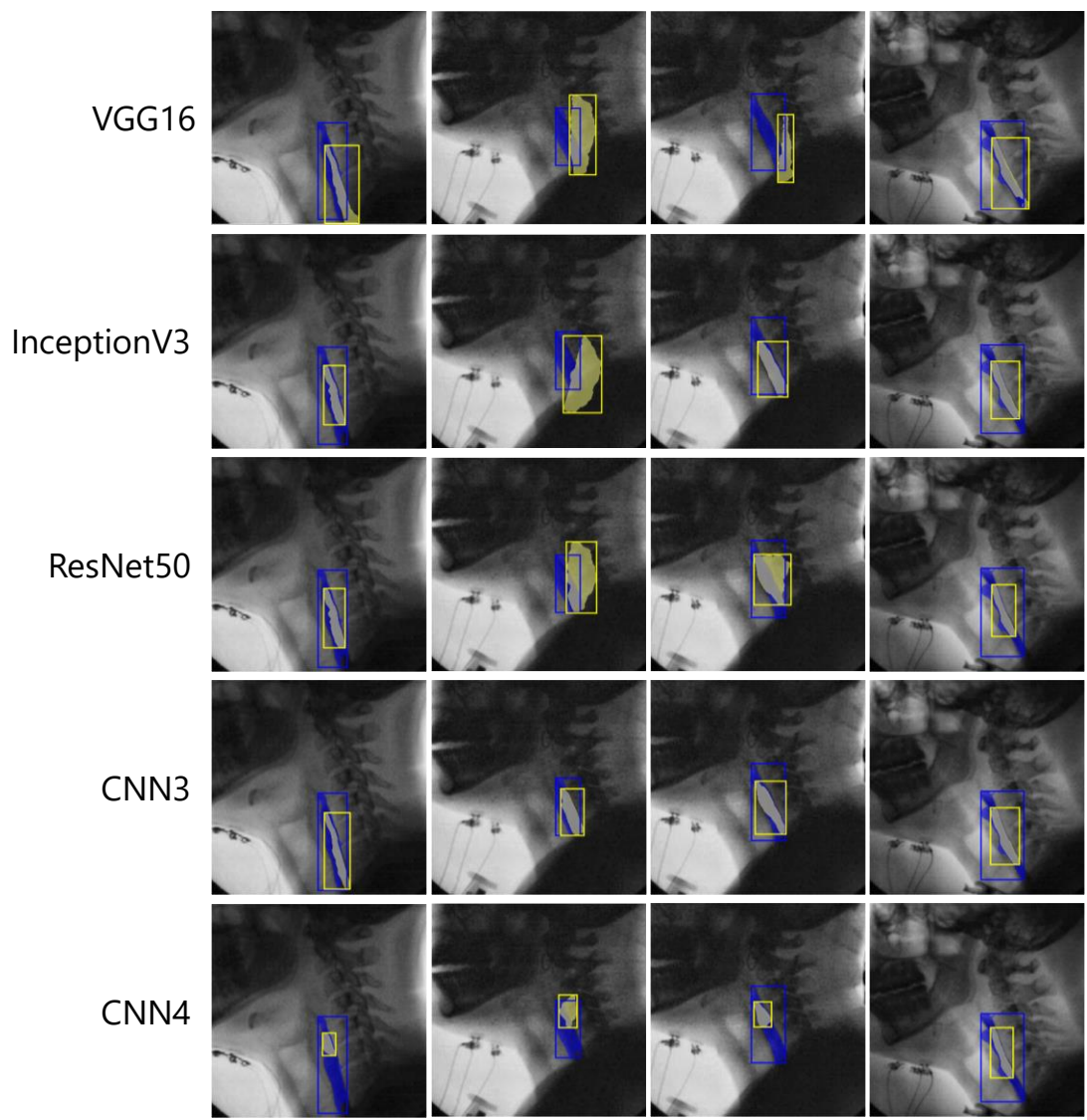

**Figure 5. Qualitative results of the weakly-supervised bolus segmentation (in yellow) obtained with the 5 CNN backbones on four sample frames. Ground truth boluses are highlighted in blue. Among the 5 CNN models, the one using the CNN3 backbone (fourth row) yielded the best results in terms of overlap between the predicted and the ground truth bolus locations.**

**Table 5. Bolus localization performance by consistency. Best results for each metric are highlighted in bold. For the RMSE we reported the median and the inter-quartile range (in parentheses).**

| | CNN backbone | $r_y$ | RMSE |
|---|---|---|---|
| Thin boluses | VGG16 | 0.340 | 0.806 (0.516-1.111) |
| | InceptionV3 | 0.744 | **0.276 (0.123-0.471)** |
| | ResNet50 | **0.945** | 0.331 (0.204-0.491) |
| | CNN3 | 0.699 | **0.256 (0.116-0.529)** |
| | CNN4 | 0.787 | 0.360 (0.235-0.623) |
| | Friedman's test | | $X^2(4) = 83.12, p < .001$ |
| Slightly thick boluses | VGG16 | 0.199 | 0.569 (0.391-0.935) |
| | InceptionV3 | 0.845 | **0.332 (0.199-0.499)** |
| | ResNet50 | **0.915** | 0.432 (0.370-0627) |
| | CNN3 | 0.725 | **0.273 (0.160-0.654)** |
| | CNN4 | 0.697 | **0.323 (0.206-0.575)** |
| | Friedman's test | | $X^2(4) = 84.59, p < .001$ |
| Mildly thick boluses | VGG16 | 0.823 | 0.574 (0.329-0.811) |
| | InceptionV3 | 0.861 | 0.251 (0.155-0.359) |
| | ResNet50 | 0.948 | 0.275 (0.157-0.348) |
| | CNN3 | **0.964** | **0.186 (0.095-0.265)** |
| | CNN4 | **0.964** | **0.221 (0.100-0.341)** |
| | Friedman's test | | $X^2(4) = 139.85, p < .001$ |
| Moderately thick boluses | VGG16 | 0.815 | 0.655 (0.440-0.985) |
| | InceptionV3 | 0.921 | 0.264 (0.147-0.474) |
| | ResNet50 | 0.939 | **0.204 (0.128-0.280)** |
| | CNN3 | 0.899 | **0.158 (0.075-0.276)** |
| | CNN4 | **0.960** | **0.180 (0.102-0.295)** |
| | Friedman's test | | $X^2(4) = 170.84, p < .001$ |
| Extremely thick boluses | VGG16 | 0.877 | 0.513 (0.317-0.755) |
| | InceptionV3 | 0.979 | 0.192 (0.119-0.342) |
| | ResNet50 | 0.963 | 0.337 (0.244-0.490) |
| | CNN3 | 0.940 | **0.132 (0.081-0.212)** |
| | CNN4 | **0.980** | **0.137 (0.073-0.264)** |
| | Friedman's test | | $X^2(4) = 178.22, p < .001$ |

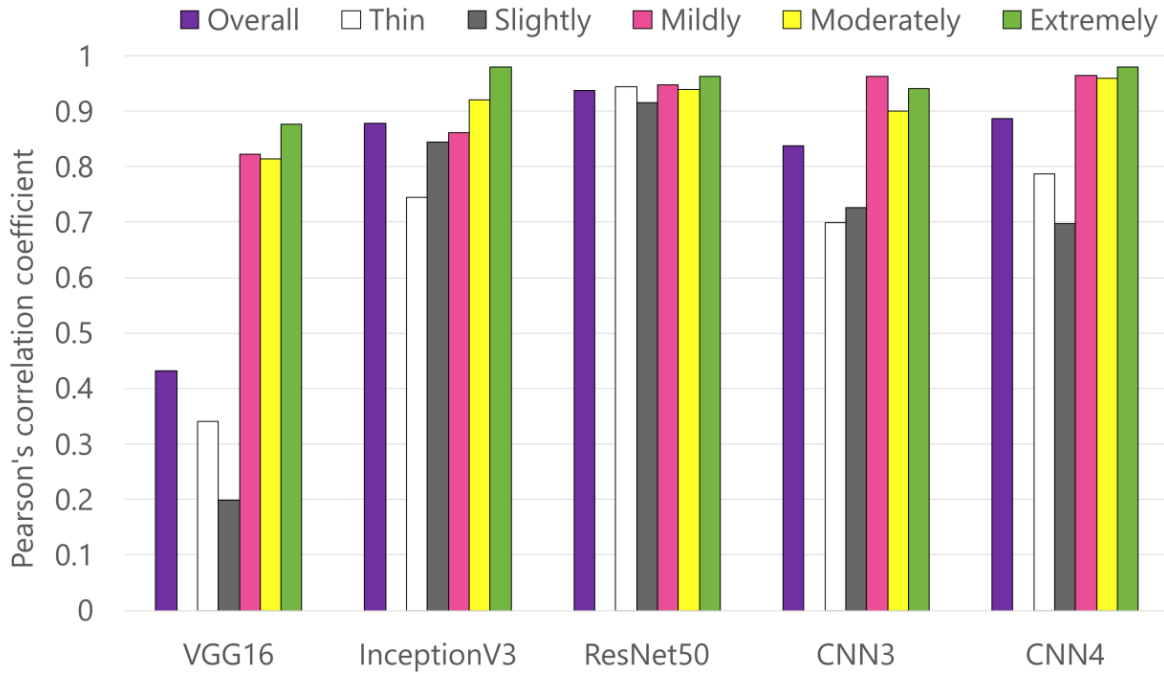

**Figure 6. Pearson's correlation coefficient of the estimated bolus trajectories with the ground truth trajectories obtained with the 5 CNN architectures for the 5 bolus consistencies.**

## 4 DISCUSSION

We have proposed a deep-learning framework for pharyngeal phase detection and bolus localization in VFSS bolus-level clips that requires only the initial and final frames of the pharyngeal phase as ground truth information for training. For the first time, we have demonstrated that pharyngeal phase detection and bolus localization are two interdependent tasks that can be tackled within the same framework. The main advantage of the proposed approach is that no manual annotation was required to develop the bolus localization algorithm. The temporal and spatial information extracted with this method can be used as input to further VFSS processing, for example, to automatically detect other timing parameters between BPM and UESC (e.g., first frame of laryngeal vestibule closure or the first frame of UES opening) [10] and to eventually improve the efficiency and accuracy of VFSS rating that is lengthy and requires extensive training.

The findings of this work are threefold and answer the following questions: 1) How far are we from implementing an automated method for detecting the pharyngeal phase in VFSS data? 2) is it possible to detect the bolus in a weakly supervised manner? and 3) is the proposed algorithmic framework robust across different bolus consistencies?

## 4.1 How far are we from implementing an automated method for pharyngeal phase detection in VFSS?

The comparison across multiple CNNs for detecting the pharyngeal phase in VFSS bolus-level clips provided us with interesting findings. Despite excellent detection performance, with F1-scores often higher than 0.9, we demonstrated that the end of the pharyngeal phase (i.e., UESC frame) could be detected more accurately than its beginning (i.e., BPM frame). Trained rater inter-rater agreement values could be easily exceeded by the automated method when detecting the UESC frames, but not for BPM frame detection. Thus, the proposed approach - specifically, 4 out of 5 CNN models - is as accurate as a trained annotator in detecting the end of the pharyngeal phase, but further improvements are needed to automatically identify the beginning of this phase.

Our results are in line with Lee et al. [15], where the pharyngeal phase could be detected with accuracy higher than 0.9 using a VGG16-based approach. In particular, the authors demonstrated that retraining all VGG16's blocks led to better performance, and for this reason we decided to train our models from scratch. Although a fair comparison was not possible due to the different datasets, we confirmed their results by demonstrating that VGG16 is able to recognize the pharyngeal phase with high performance. Moreover, our results went further, and demonstrated that other CNN architectures, such as ResNet50, can provide better results.

To foster the clinical adoption of this method, in addition to improving the BPM frame detection, we will need further validation on larger datasets composed of participants with swallowing disorders and recorded with different workstations. This will allow us to fine-tune the proposed approach and conduct domain adaptation experiments to develop algorithms tailored to the specific clinical application and setting.

## 4.2 Weakly-supervised bolus localization

The results obtained with our weakly-supervised approach for localizing the bolus are promising, with correlations with the ground truth trajectories higher than 0.9. To the best of our knowledge, this is the first time that bolus localization in VFSS has been tackled in a weakly-supervised manner. One of the main contributions of this manuscript is having demonstrated that an accurate pharyngeal phase detector can also be used to localize the bolus. This approach opens new scenarios for automated bolus detection in VFSS, for instance, the development of novel algorithms for unsupervised bolus tracking or hybrid methods that combine unsupervised and supervised segmentations to boost performance. A specific application will be the use of the coarse bolus localization results as attention maps to guide the region proposal selection of supervised segmentation approaches, such as the one based on Faster R-CNN proposed by Caliskan et al. [16].

Our results showed that there may be a bias in localizing the bolus, for example with ResNet50, where correlations were constantly higher than 0.9 (see Tables 4 and 5) but RMSE and F1-scores were inferior to other architectures. This could be due to the changing shape of the bolus that becomes elongated during the pharyngeal phase; thus, the algorithm was able to detect only a portion of the entire bolus, and this would justify the reduction of F1-score at higher IoU thresholds (Figure 4).

From the results we also observed that not all models with good performance on pharyngeal phase detection showed equivalently good performance on the bolus localization task. The most striking example is VGG16, which was one of the top-two models for pharyngeal phase detection but was the worst for localizing the bolus (Figures 4-6, Tables 4 and 5). The poor localization performance can be attributed to the smaller heatmaps that presented wider and smoothed attention regions, thereby highlighting the area of the bolus in a non-precise way and skewing the rest of the bolus localization algorithm. Surprisingly, CNN3 and CNN4 provided the best results in terms of bolus localization, especially for thicker boluses. This behaviour can be explained by the larger size of the original activation maps, which can provide more detailed initializations to the active contours. Thus, an interesting development of this work will be finding the best trade-off between the CNN layer from which the attention maps are extracted and the bolus localization performance.

### 4.3 Robustness of the proposed approach with respect to consistencies

To the best of our knowledge, this is the first time that pharyngeal phase detection and bolus localization performance have been calculated for different consistencies. For detection of the BPM frame, it is possible to notice a decrease in performance with increasing bolus thickness. This phenomenon may be explained by the presence of pre-existing residue in the oropharynx, which interferes with accurate detection by the network of new BPM events [48]. Conversely, UESC frames can be detected with error below 3 frames in more than 90% of the cases, without a clear trend across consistencies. The extension of this analysis to other types of swallows (e.g., piecemeal swallows) can make the approach more robust to the presence of smaller amounts of material, thus improving performance in BPM detection.

A trend in the performance across different consistencies was also visible in the bolus localization results. However, unlike the BPM detection trend, in this case, the correlation coefficients improved for thicker consistencies. The opposite trend can be seen with the RMSE values, which decreased with thicker consistencies. This behaviour can be explained by the fact that thinner boluses are faster and less defined in shape than thicker ones. From Figure 6, we can detect a cut-off in correspondence to mildly thick boluses, after which the performance increased for all models except ResNet50, which produced robust performance across all consistencies. However, the relatively higher RMSE values and lower F1-scores, suggested the presence of a bias in the bolus localization results obtained with ResNet50.

### 4.4 Limitations and future work

In this work we included only healthy participants and single swallow clips. Future work will address this problem in participants with neurological diseases and swallowing impairment, for instance by using domain adaptation techniques, to transfer the knowledge learned on this dataset to another data distribution. Another limitation was that we focused only on the first swallow event of the bolus-level clips, without considering the presence of residue or piecemeal swallows. In particular, the detection of residue is an important task in VFSS rating, as residues in the oropharynx are associated with a higher risk of aspiration-penetration [48]. Although we demonstrated the feasibility of using the class activation maps for localizing the bolus, further work is needed to refine the bolus segmentation steps, to be able to also locate the presence of residues. Considering that we retained only the largest blob from the binarized maps, it is not possible in the current form of the algorithm to locate the presence of residues when the main body of the bolus is visible within the pharynx.

The expansion of the dataset in terms of annotated frames will allow us retraining and testing additional bolus segmentation algorithms [16], [49], to conduct a fair comparison among multiple approaches for bolus localization, and refine the weakly-supervised bolus detection method. In fact, part of the proposed algorithm (in particular the image processing steps implemented for refining the bolus maps) will benefit from hyperparameter tuning, for example by detecting the optimal threshold for binarizing the maps, and the optimal approaches for segmenting the blobs.

With further improvements and validation of the proposed algorithmic framework, we are confident that our approach can be used to automatically extract clinical parameters of interest from bolus-level clips. Given the modular nature of the proposed framework, we will also test additional CNN architectures and implement recurrent layers (e.g., gated recurrent unit and long short-term memory [50]) to improve the temporal detection of the BPM frame.

# 5 CONCLUSION

We have presented a deep-learning framework for VFSS analysis that jointly tackles pharyngeal phase detection and bolus localization in a weakly-supervised manner, by exploiting the strong correlations between the class-activation maps and the bolus presence in the pharynx. For the first time, we have demonstrated that these two tasks can be solved jointly with the same model and without any manual annotations of bolus location. As far as pharyngeal phase detection is concerned, we obtained performance close to the gold standard (i.e., trained clinical raters). However, to envision a translation of this approach into clinical practice, further experiments are needed to detect the beginning of the pharyngeal phase and to localize the bolus in a robust manner across all bolus consistencies. With further validation on expanded datasets that include data from individuals with swallowing impairment, we are confident that our approach will support clinicians during VFSS rating, with the end goal of providing objective clinical assessment and optimizing treatment planning for individuals with dysphagia.

# 6 ACKNOWLEDGMENTS

This work was supported by an R01 grant from the National Institute on Deafness and Other Communication Disorders (Grant DC011020) to the last author. The authors would like to thank